\documentclass[9pt,journal]{IEEEtran}
\usepackage{color}
\usepackage{dsfont}
\usepackage{multicol}
\usepackage{mathrsfs}
\usepackage{amsfonts}
\usepackage{latexsym}
\usepackage{graphicx}
\usepackage{amsmath}
\usepackage{epstopdf}
\usepackage{subfig}
\usepackage{cite}
\usepackage{enumerate}
\usepackage{amssymb}
\usepackage{bm}
\newtheorem{theorem}{\textbf{Theorem}}
\newtheorem{lemma}{\textbf{Lemma}}
\newtheorem{remark}{\textbf{Remark}}
\newtheorem{definition}{\textbf{Definition}}
\newtheorem{assumption}{\textbf{Assumption}}

\newcommand*{\QEDB}{\hfill\ensuremath{\square}}

\def\BibTeX{{\rm B\kern-.05em{\sc i\kern-.025em b}\kern-.08em
    T\kern-.1667em\lower.7ex\hbox{E}\kern-.125emX}}
\begin{document}

\title{{\huge{Asymptotic Tracking Control of Uncertain MIMO Nonlinear Systems with Less Conservative Controllability Conditions}}}
\author{Bing Zhou,  Xiucai Huang, and Yongduan Song
        % <-this % stops a space
 \thanks{Bing Zhou, Xiucai Huang and Yongduan Song are with the Chongqing Key Laboratory of Intelligent Unmanned Systems, School of Automation, Chongqing University, Chongqing, 400044,  China (e-mail: bingbing\_z@foxmail.com; hxiucai@cqu.edu.cn; ydsong@cqu.edu.cn)
}}

\markboth{}
{Shell \MakeLowercase{\textit{et al.}}: Bare Demo of IEEEtran.cls for Journals}

\maketitle

\begin{abstract}
For uncertain multiple inputs multi-outputs (MIMO) nonlinear systems, it is nontrivial to achieve asymptotic tracking, and most existing methods normally demand certain controllability conditions that are rather restrictive or even impractical if unexpected actuator faults are involved. In this note, we present a method capable of achieving zero-error steady-state tracking with less conservative (more practical) controllability condition. By incorporating a novel Nussbaum gain technique and some positive integrable function into the control design, we develop a robust adaptive asymptotic tracking control scheme for the system with time-varying control gain being unknown its magnitude and direction. By resorting to the existence of some feasible auxiliary matrix, the current state-of-art controllability condition is further relaxed, which enlarges the class of systems that can be considered in the proposed control scheme. All the closed-loop signals are ensured to be globally ultimately uniformly bounded. Moreover, such control methodology is further extended to the case involving intermittent actuator faults, with application to robotic systems. Finally, simulation studies are carried out to demonstrate the effectiveness and flexibility of this method.
\end{abstract}

% Note that keywords are not normally used for peer review papers.

\IEEEpeerreviewmaketitle

\section{Introduction}
In this article, we consider a class of $n$th order MIMO uncertain nonlinear systems in canonical form \cite{Bechlioulis2017tac}
\begin{equation}
\begin{cases}
\label{os}
\dot{\bm{x}}_{i}=\bm{x}_{i+1},\quad i=1,2,\ldots,n-1, \\
\dot{\bm{x}}_{n}=\bm{f}(\bar{\bm{x}})+\bm{g}(\bar{\bm{x}},t)\bm{u}_{a}+{\bm{d}}(\bar{\bm{x}},t),\\
\bm{y}=\bm{x}_1,
\end{cases}
\end{equation}
where $\bm{x}_{i}=[x_{i1},\ldots,x_{im}] \in \mathbb{R}^{m}$ $ (i=1,\ldots,n)$ is the $i$th state vector;
$\bar{\bm{x}} = [\bm{x}_1^T,\ldots,\bm{x}_n^T]^T \in \mathbb{R}^{m\times n}$;
$\bm{u}_{a}=[u_{a1},\ldots,u_{am}]^T \in \mathbb{R}^m$ is the actual input vector of the system (the output of actuators);
${\bm{d}}(\bar{\bm{x}},t)\in \mathbb{R}^m$ denotes the lumped uncertainties and external disturbances;
${\bm{y}}= [y_1, \ldots , y_m]^T \in \mathbb{R}^m$ is the output vector;
${\bm{f}}(\bar{\bm{x}})\in \mathbb{R}^m$ is a smooth but unknown nonlinear function vector;
${\bm{g}}(\bar{\bm{x}},t) \in \mathbb{R}^{m \times m}$ is the input gain matrix and its magnitude and sign are both unknown.
{Before feeding the control into such uncertain nonlinear systems, the foremost task is to prevent the loss of controllability. Such task is a hard nut if the input gain is unknown. For single-input single-output (SISO) systems, the input gain is commonly assumed to be bounded away from the origin to ensure the controllability of the systems \cite{Song-Wang2017auto,Huang-Song2020auto}. Such condition can be readily extended to the case with partial loss of effectiveness (PLOE) actuator faults since there is no coupling between the input gain and the coefficient of  faults \cite{Zhang-Yang2017tac,Yang2021jcd}. While for MIMO uncertain nonlinear system (\ref{os}), such problem is far more involved due to the intrinsic coupling among different control channels, making the tracking control for such kind of systems much more challenging yet interesting.}

In order to ensure the controllability of system (\ref{os}), the primitive corresponding condition refers to that the minimum singular value of $\bm g$ has a positive lower bound \cite{Liu1993ijc} and the norm of $\bm g$ has a positive upper bound \cite{Chan2001tac}.
Thereafter, several efforts have been made to relax the restrictions on such controllability condition.
The first one was reported in \cite{Xu2003tac} by assuming that $\bm g+\bm {g}^T$ is uniformly positive (or negative) definite, leading to consequent results on tracking control for uncertain nonlinear MIMO systems (see \cite{Bechlioulis2008tac,Katsoukis2021tac,Song-Huang2016tac,Theodorakopoulos2016tac,Bikas2021tac} for examples).
However, such condition is still restrictive since some controllable scenarios are overlooked.
A further controllability relaxation was made in \cite{Lee2016scl} by importing an auxiliary matrix, yet such auxiliary matrix is required to be known. Besides, all the controllability conditions in \cite{Liu1993ijc,Chan2001tac,Xu2003tac,Bechlioulis2008tac,Song-Huang2016tac,Katsoukis2021tac,Theodorakopoulos2016tac,Bikas2021tac} could be invalid once the actuators suffer from the PLOE.
Different from the SISO case \cite{Zhang-Yang2017tac,Yang2021jcd}, the fault-tolerant control (FTC) results for uncertain MIMO nonlinear systems are relatively rare due to the challenging arising from the coupling between $\bm{g}$ and the faults coefficient matrix $\bm{\rho}$.
 To address this problem, a controllability condition is established in \cite{Jin2019tcyb} by assuming that the positive/negative definite of $\bm {g}\bm{\rho}+\bm{\rho}\bm {g}^T$, which, however, could leave out some controllability scenarios.
 Although an upgraded condition was considered in \cite{Zhang-Yang2020auto}, it is only valid for Euler-Lagrange (EL) systems and cannot handle intermittent actuator faults.
 More recently, another controllability relaxation was implemented in \cite{Huang-Song2021tac}, where the authors coined a new controllability condition by resorting to an unknown auxiliary matrix, that includes the conditions in \cite{Lee2016scl,Jin2019tcyb,Zhang-Yang2020auto} as special ones, yet the matrix is limited to a diagonal form.
 {Therefore, it remains to be an interesting yet challenging problem to relax the controllability condition for uncertain MIMO nonlinear systems in the absence/presence of multiplicative intermittent actuator faults.}

{Note that all the above controllability conditions are based on the fact that the control direction (the sign of the control coefficient or high frequency gain) of the system is known, which further limits their application.
Whereas, the adaptive tracking control design under unknown control direction is quite difficult and the Nussbaum functions invented in  \cite{Nussbaum1983scl} are normally adopted.
At present, there have full-fledged results in SISO systems \cite{Ye1998tac,Ge2003tac,Ge2004smc} and MIMO systems \cite{Jiang2006auto} for the case that the control coefficient is a constant of unknown sign or value.
However, for the case of time-varying control coefficients with unknown sign and value, the results have been obtained only for some particular Nussbaum functions \cite{Liu2006tac,Bechlioulis2009auto,Shi2014tfs,WangC-Wen2021tac} and rarely achieve global asymptotic convergence, notably for uncertain MIMO nonlinear systems \cite{Shi2014tfs}. As far as we know, with less conservative controllability conditions, the adaptive asymptotic tracking control problem of (\ref{os}) under unknown control gain and control direction remains unsolved.}

{Motivated by the observation above, this paper investigates the tracking control problem for a class of unknown nonlinear MIMO systems under certain milder controllability condition, two sets of robust and adaptive control schemes are proposed for the cases in the absence and presence of intermittent actuator faults, respectively.
The features and contributions of the developed schemes are summarized as follows.}
\begin{enumerate}[\indent 1)]
\item {By assuming the existence of some feasible auxiliary matrices, the controllability conditions reported in the relevant literatures \cite{Bechlioulis2008tac,Song-Huang2016tac,Katsoukis2021tac,Theodorakopoulos2016tac,Bikas2021tac,Lee2016scl} are further relaxed for uncertain MIMO nonlinear systems, which is also extended to the case with unexpected actuator faults, making the controllability conditions more practical than those in \cite{Zhang-Yang2020auto,Cao2020ijc,Jin2019tcyb,Huang-Song2021tac} and thus enlarges the class of systems that can be considered.}
\item {In contrast to \cite{Wang-Wen2010auto,WangW-Wen2017auto,Bechlioulis2008tac},
    the proposed method elegantly handles the unknown nonlinearities by extracting the readily computable deep-rooted information without using any linearization and approximation mechanisms (despite the relaxation of the controllability condition), making the developed control scheme simpler in structure and less-expensive in computation.
    Furthermore, the unexpected actuator faults can be compensated automatically without resorting to any additional fault detection and diagnosis module.}
\item {By incorporating a novel Nussbaum gain technique from \cite{ChenZY2019auto} and some positive integrable function from \cite{WangW-Wen2017auto} into the control design, global asymptotic tracking control is achieved with the control gain and its direction being completely unknown, and the ultimate uniform boundedness of all the internal signals is guaranteed.}

\end{enumerate}
In addition, the proposed method is further applied to robotic systems with unexpected intermittent actuator faults. Simulation results are provided to verify its efficacy.

%\emph{Organization:}
The remainder of this paper is organized as below. In Section II, some preliminaries and the control problem are introduced.
The control design and the stability analysis for the cases in the absence/presence actuator faults are presented in Sections III and IV, respectively.
In Section V, the proposed control scheme is applied to the practical robotic systems with simulation verification provided.
Finally, the conclusions are given in section VI.

\section{Preliminaries and Problem Statement }
\subsection{Notation}
We use bold notations to denote matrices (or vectors).
For a nonsingular matrix $\bm{A}\in \mathbb{R}^{m \times m}$, $\textrm{min}\{\textrm{eig}(\bm{A})\}$, $\textrm{max}\{\textrm{eig}(\bm{A})\}$ and $\textrm{min}\{\textrm{sv}(\bm{A})\}$ calculate the minimum eigenvalue,  maximum eigenvalue and minimum singular value of $\bm A$, respectively. $\bm{I}_m \in \mathbb{R}^{m \times m}$ is the identity matrix.
$\bm{0}_m \in \mathbb{R}^m$ stands for a vector of zeros. $\|\bm{\cdot}\|$ denotes the standard Euclidean norm.
\subsection{Nussbaum Function}
To deal with the unknown control directions associated with the time-varying control gain $\bm{g}(\bar{\bm{x}},t)$ in (\ref{os}), some special Nussbaum-type functions should be introduced.
\begin{definition}\cite{ChenZY2019auto}
{A continuously differentiable function $\hbar(\zeta) : [0, \infty)\rightarrow(-\infty, \infty)$ is called a $BL$-type Nussbaum function, if for any constant $L > 1$, it satisfies
  \begin{subequations}
  \begin{align}
   &\lim\limits_{\zeta \to \infty}\frac{\int_{0}^{\zeta}\hbar^{+}(\tau)d\tau}{\zeta}=\infty, \ \limsup\limits_{\zeta \to \infty}\frac{\int_{0}^{\zeta}\hbar^{+}(\tau)d\tau}{\int_{0}^{\zeta}\hbar^{-}(\tau)d\tau}\geq L,\label{blA}\\
     & \lim\limits_{\zeta \to \infty}\frac{\int_{0}^{\zeta}\hbar^{-}(\tau)d\tau}{\zeta}=\infty, \ \limsup\limits_{\zeta \to \infty}\frac{\int_{0}^{\zeta}\hbar^{-}(\tau)d\tau}{\int_{0}^{\zeta}\hbar^{+}(\tau)d\tau}\geq L \label{blB}
  \end{align}
  \end{subequations}}
\end{definition}
 {with $\hbar^{+}(\zeta)= \textrm{max}\{0,\hbar(\zeta)\}$ and $\hbar^{-}(\zeta)= \textrm{max}\{0,-\hbar(\zeta)\}$. Particularly, if $L = \infty$, the $BL$-type Nussbaum function can be simply called a $B$-type Nussbaum function.
For example, $\zeta^2\sin(\zeta)$ and $\exp(\zeta)\sin(\zeta)$ are $BL$-type Nussbaum functions; $\exp(\zeta^2)\sin(\zeta) $ and $\exp(\zeta^2)\cos((\pi/2)\zeta)$ are $B$-type Nussbaum functions. For later analysis, we need the following corresponding lemma.}

\begin{lemma}\cite{ChenZY2019auto}\label{lemma0}
{Consider two continuously differentiable functions $V(t) : [0,\infty)\rightarrow \mathbb{R}^+$, $\zeta(t) : [0,\infty)\rightarrow \mathbb{R}^+$. Let $\psi(t) : [0,\infty) \rightarrow [\underline{\psi},\bar{\psi}]$ with $\underline{\psi}$ and $\bar{\psi}$ satisfying $\underline{\psi}\bar{\psi}>0$. For some constant $\iota$ and $BL$-type Nussbaum function $\hbar(\zeta(t))$ with $L > \textrm{max}\{({\underline{\psi}}/{\bar{\psi}}),({\bar{\psi}}/{\underline{\psi}})\}$, if it holds that
\begin{equation}
\begin{cases}
\label{LBS}
  \dot{V}(t)\leq\big(\psi(t)\hbar(\zeta(t))+\iota\big)\dot{\zeta}(t), \\
  \dot{\zeta}(t) \geq0,\ \forall t\geq0,
\end{cases}
\end{equation}
then $V(t)$ and $\zeta(t)$ are bounded over $[0,\infty)$. In particular, the statement holds for $L = \infty$.}
\end{lemma}

\begin{remark}\label{Remark1}
{Note that although Nussbaum gain techniques have been widely used in dealing with time-varying control coefficients with unknown signs (see \cite{Ge2004smc,Jiang2006auto,Liu2006tac,Bechlioulis2009auto,Shi2014tfs} for examples), those results could only be valid during some finite time interval, not for all $t \in [0,\infty)$. Recently, Chen coined a class of \emph{BL}-type Nussbaum functions in \cite{ChenZY2019auto} that have been proven to be effective in tackling unknown control direction problem with time-varying input coefficients. By following this route, such special Nussbaum functions are subtly applied in our control design under certain relaxed controllability condition, as seen later.}
\end{remark}

\subsection{Problem Statement}
Consider the uncertain nonlinear MIMO system (\ref{os}).
Denote the desired trajectory as $\bm{y}^*=[y^*_1,\ldots,y^*_m] \in \mathbb{R}^m$, and define the tracking error as $\bm{e}=\bm{x}_1-\bm{y}^*$.
The control objective in this paper is to design a control law such that
\begin{enumerate}
  \item All signals in the closed-loop systems are guaranteed to be globally ultimately uniformly bounded;
  \item The output $\bm{y}$ tracks the reference signal $\bm{y}^*$ asymptotically.
  \end{enumerate}

To achieve this objective, the following assumptions and lemmas are needed.
\begin{assumption}\label{assum3}
  The desired tracking trajectory $\bm{y}^*$ and its derivatives up to $(n-1)$th order are known and bounded, and its $n$th order derivative $\bm{y}^{*(n)}$ is bounded by an unknown constant, i.e., $\|\bm{y}^{*(n)}\|\leq \bar{y}<\infty$. The system state vector $\bm{x}_i$ is available for control design.
\end{assumption}

\begin{remark}\label{Remark2}
{The first part of \emph{Assumption} \ref{assum3} is quite standard and commonly imposed in the existing works, see \cite{Bechlioulis2009auto,Song-Wang2017tac,Zhao-Song2019auto} for examples. The second part of the assumption requires that the system state are all available for control design, if not, state observer must be constructed, yet is beyond the scope of this paper.}
\end{remark}

\begin{assumption}\label{assum1}
{For uncertain system (\ref{os}), there exists an unknown symmetric and positive definite matrix $\bm{\alpha}(\bar{\bm{x}}_{n-1},t) \in \mathbb{R}^{m \times m}$ which is differentiable with respect to $t$ and ${\bm{x}}_j$ ($j=1,\ldots,n-1$), such that ${{\bm{\alpha} \bm{g}+\bm{g}^T\bm{\alpha}}}$ is either uniformly positive or uniformly negative definite, i.e., $\lambda_m\lambda_M>0$ with $\lambda_m(t):=\textrm{min}\{\textrm{eig}({{\bm{\alpha} \bm{g}+\bm{g}^T\bm{\alpha}}})\}$ and $\lambda_M(t):=\textrm{max}\{\textrm{eig}({{\bm{\alpha} \bm{g}+\bm{g}^T\bm{\alpha}}})\}$.}
\end{assumption}

\begin{assumption}\label{assum2}
  There exist some unknown positive constants $a_f$, $a_i$, $i=1,2$ and known nonnegative scalar functions $\varphi_f (\bar{\bm{x}})$, $\varphi_1(\bar{\bm{x}}_{n-1})$ and $\varphi_2(\bar{\bm{x}})$, such that
  \begin{equation}\label{afphi}
    \|\bm{f}(\bar{\bm{x}})+\bm{d}(\bar{\bm{x}},t)\|\leq a_f \varphi_f(\bar{\bm{x}}),
  \end{equation}
  \begin{equation}\label{aphi}
    \|\bm{\alpha}(\bar{\bm{x}}_{n-1},t)\|\leq a_1\varphi_1(\bar{\bm{x}}_{n-1}),\,\big\|\frac{\partial \bm{\alpha}}{\partial t}\big\|\leq a_2\varphi_2(\bar{\bm{x}}),
  \end{equation}
  where $\varphi_f(\bar{\bm{x}})$, $\varphi_1(\bar{\bm{x}}_{n-1})$ and $\varphi_2(\bar{\bm{x}})$ are radially unbounded.
\end{assumption}

\begin{remark}\label{Remark3}
{\emph{Assumption} \ref{assum1} ensures the controllability of system (\ref{os}), which is milder than the current state-of-art \cite{Bechlioulis2008tac,Song-Huang2016tac,Katsoukis2021tac,Theodorakopoulos2016tac,Bikas2021tac,Lee2016scl}.
  Specifically, when $\bm{\alpha} = \bm{I}_m$ and the control direction is known, \emph{Assumption} \ref{assum1} corresponds to the traditional controllability conditions in \cite{Bechlioulis2008tac,Song-Huang2016tac,Katsoukis2021tac,Theodorakopoulos2016tac,Bikas2021tac},
  which broadcast that $\bm{g}+\bm{g}^T$ is either uniformly negative definite or uniformly positive definite.
  Unfortunately, such assumption is restrictive more or less.
  For example, when
  \begin{equation}\label{exg1}
  \bm{g}({\bm{\bar{x}}}_{2},t)=\begin{bmatrix}
    2 & 0.6+0.1\cos(x_{11}x_{21}) \\
    2+0.1\cos(x_{11}x_{21}) & 0.9
  \end{bmatrix},
  \end{equation}
   it is readily verified that $\bm{g}+\bm{g}^T$ is neither uniformly positive nor uniformly negative definite when $\cos(x_{11}x_{21})>0.42$.
   However, such scenario is still within the scope of \emph{Assumption} \ref{assum1} for $x_{11}, x_{21}\in R$, once choosing
  \begin{equation}\label{exa1}
  \bm{\alpha}({\bm{x}}_{1},t)=\begin{bmatrix}
    0.9+0.1\sin(t) & 0.1\sin(x_{11}) \\
    0.1\sin(x_{11}) & 0.4+0.1\cos(t)
  \end{bmatrix},\  t\in[0,\infty).
  \end{equation}
  Although similar assumption is illustrated in \cite{Lee2016scl}, $\bm{\alpha}$ therein is required to be known.}
  {For these reasons, the controllability conditions in \cite{Bechlioulis2008tac,Song-Huang2016tac,Katsoukis2021tac,Theodorakopoulos2016tac,Bikas2021tac,Lee2016scl} can be viewed as some special cases of ours since both $\bm{\alpha}$ and the control direction are allowed to be unknown in our case.}
\end{remark}

\begin{remark}\label{Remark4}
{It is worth noting that in \emph{Assumption} \ref{assum1} only the existence of the auxiliary matrix $\bm{\alpha}$ is required, while its analytically identification is not needed. This is because such $\bm{\alpha}$ is only used for stability analysis but not involved in control design.  Interestingly, the existence of $\bm{\alpha}$ is naturally satisfied in many practical systems, such as robotic systems \cite{Cao2020ijc}, wheeled inverted pendulum systems \cite{Li2010auto}, unmanned aerial vehicle (UAV) systems \cite{Chen2016tie}, and high speed train (HST) systems \cite{Song-Yuan2016tie}, therein the first two ones are essentially in the form of EL models, thus $\bm{\alpha}$ can be immediately chosen as $\bm{\alpha} = \bm{I}_m$ \cite{Bechlioulis2008tac,Song-Huang2016tac,Katsoukis2021tac,Theodorakopoulos2016tac,Bikas2021tac}, while for UAV and HST systems, the controllability condition in \emph{Assumption} \ref{assum1} is still valid by trivially choosing $\bm{\alpha}$ \cite{Chen2016tie,Song-Yuan2016tie}.}
\end{remark}

\begin{remark}\label{Remark5}
  {Since $\bm{\alpha}(\bar{\bm{x}}_{n-1},t)$ is a time-varying and unknown nonlinear function, its introduction will inevitably bring additional difficulties in both control design and stability analysis. In order to address those obstacles, certain conditions (\ref{afphi}) and (\ref{aphi}) are imposed in \emph{Assumption} \ref{assum2}, respectively, which are related to the extraction of the deep-rooted information from the model. Such task can be readily fulfilled for any practical system with only crude model information \cite{Song-Huang2016tac}.
  In such a way, the local Lipschitz conditions for $\bm{f}$, $\bm{g}$, ${\bm{d}}$ and $\bm{\alpha}$ required in \cite{Kanakis2020tac,Huang-Song2021tac} are no longer needed anymore. Particularly, even if no any prior information about $\bm{\alpha}$ can be extracted, one can judiciously choose $\varphi_1=\varphi_2=1$.
  In such case, it is  equivalent to impose some upper bounds on $\|\bm{\alpha}\|$ and $\|\partial{\bm{\alpha}}/\partial t\|$, which however, is still more general than the conditions imposed in \cite{Bechlioulis2008tac,Song-Huang2016tac,Katsoukis2021tac,Theodorakopoulos2016tac,Bikas2021tac,Lee2016scl}.}
\end{remark}

\begin{lemma}\label{lemma1}
  For any $\bm{\phi} \in \mathbb{R}^m$ and any $\nu(t)\geq0$, it holds that
  \begin{subequations}
  \begin{align}
    \|\bm{\phi}\|\leq &\frac{\|\bm{\phi}\|^2}{\|\bm{\phi}\|+\nu(t)}+\nu(t), \label{ZA}\\
    \|\bm{\phi}\|\leq & \frac{\|\bm{\phi}\|^2}{\sqrt{\|\bm{\phi}\|^2+\nu(t)^2}}+\nu(t) \label{ZB}.
  \end{align}
 \end{subequations}
\end{lemma}

\emph{\textbf{Proof:}}\ Since $\nu(t)\geq0$, we have
  \begin{equation}
  \begin{split}
    \|\bm{\phi}\|=&\frac{\|\bm{\phi}\|^2 + \|\bm{\phi}\|\nu(t))}{\|\bm{\phi}\|+\nu(t)}
    \leq \frac{\|\bm{\phi}\|^2}{\|\bm{\phi}\|+\nu(t)}+\nu(t),
    \end{split}\nonumber
  \end{equation}
and
\begin{equation}
  \begin{split}
    \|\bm{\phi}\|=&\frac{\|\bm{\phi}\| \sqrt{\|\bm{\phi}\|^2+\nu(t)^2}}{\sqrt{\|\bm{\phi}\|^2+\nu(t)^2}}\leq\frac{\|\bm{\phi}\|^2}{\sqrt{\|\bm{\phi}\|^2+\nu(t)^2}}+\nu(t).
    \end{split}\nonumber
  \end{equation}
  The proof is completed.\QEDB
\begin{remark}\label{Remark6}
It is interesting to note that the inequality (\ref{ZB}) is widely used in \cite{WangW-Wen2017auto,WangC-Wen2021tac}, yet it is a little conservative than inequality (\ref{ZA}) since
  \begin{equation}
  \begin{split}
    \frac{\|\bm{\phi}\|^2 }{\|\bm{\phi}\|+\nu(t)}+\nu(t)\leq\frac{\|\bm{\phi}\|^2}{\sqrt{\|\bm{\phi}\|^2+\nu(t)^2}}+\nu(t).
    \end{split}
  \end{equation}
{The difference lies in the right hand side (RHS) of (\ref{ZB}) is differentiable with respect to time while the RHS of (\ref{ZA}) does not need to be differentiable. In this work, the inequality (\ref{ZA}) will be applied since the derivative of the RHS of (\ref{ZA}) is not needed in the stability analysis as seen later.}
\end{remark}

\begin{lemma}\cite{Horn1990book}\label{lemma5}
  Let $\bm{\Lambda}$ be an $n\times n$ symmetric matrix and $\underline{\bm{x}}$ be a nonzero vector, if $\beta=\frac{\underline{\bm{x}}^T\bm{\Lambda}\underline{\bm{x}}}{\underline{\bm{x}}^T\underline{\bm{x}}}$, then there is at least one eigenvalue of $\bm{\Lambda}$ in the interval $(-\infty,0]$ and at least one in $[0,\infty)$.
\end{lemma}

\section{Control Without Actuator Fault}
{In this section we develop the tracking controller with controllability relaxation for uncertain MIMO nonlinear system (\ref{os}) without actuator fault.}
To facilitate the later technical development, we first introduce a filtered variable $\bm{s}(t)$ as follows:
\begin{equation}
\label{z}
\bm{s}(t)=\lambda_{1}\bm{e}_{1}+\lambda_{2}\bm{e}_{2}+\cdots+\lambda_{n-1}\bm{e}_{n-1}+\bm{e}_{n}
\end{equation}
with $\bm{e}_1=\bm{e}$ and $\dot{\bm{e}}_i=\bm{e}_{i+1} (i=1,\ldots,n-1)$. Let $\lambda_i (i=1,\ldots,n-1)$ be some constants chosen properly by the designer such that the polynomial $\bm{z}^{n-1}+\lambda_{n-1}\bm{z}^{n-2}+\cdots+\lambda_{1}$ is Hurwitz.
Then the following lemma is introduced.
\begin{lemma}\label{lemma2}\cite{Song-Wang2017tac}
  Consider the filtered error $\bm{s}(t)$ given in (\ref{z}), if $\bm{s} \rightarrow \bm{0}_m$ as $t \rightarrow \infty$, then the original error $\bm{e}(t)$ and its derivative up to $n$th order converge asymptotically to zero as $t\rightarrow \infty$ with the same decreasing rate as that of $\bm{s}(t)$.
\end{lemma}

Using (\ref{os}) and (\ref{z}), the derivative of $\bm{s}(t)$ is derived as
\begin{equation}
\begin{aligned}
\label{ds}
\dot{\bm{s}}(t)=&\,\lambda_{1}\bm{e}_{2}+\lambda_{2}\bm{e}_{3}+\cdots+\lambda_{n-1}\bm{e}_{n}-\bm{y}^{*(n)}\\
&+\bm{f}+\bm{g}\bm{u}_{a}+\bm{d}\\
=&\ \bm{\Phi}+\bm{g}\bm{u}_{a}+\bm{f}+\bm{d}- \bm{y}^{*(n)}
\end{aligned}
\end{equation}
where $\bm{\Phi}=\lambda_{1}\bm{e}_{2}+\cdots+\lambda_{n-1}\bm{e}_{n}$ is computable. From \emph{Assumptions} \ref{assum3} and \ref{assum2}, it is readily verified that
\begin{equation}
\begin{aligned}\label{L1}
\|\bm{\Phi}+\bm{f}+\bm{d}- \bm{y}^{*(n)}\| \leq \|\bm{\Phi}\| + a_f \varphi_f(\bar{\bm{x}})+\bar{y}.
\end{aligned}
\end{equation}

At this stage, we construct the control scheme as:
\begin{equation}
\begin{aligned}\label{u1}
\bm{u}_a&=\hbar(\zeta)\bm{\eta}, \\
\bm{\eta}&=k\bm{s}+\frac{{\hat\theta\varphi^2\bm{s}}}{\varphi\|\bm{s}\|+\nu(t)}
\end{aligned}
\end{equation}
with
\begin{equation}
\begin{aligned}\label{a1}
\dot{\zeta}&=\sigma_1\bm{s}^T\bm{\eta},\\
\dot{\hat{\theta}}&=\frac{{\sigma_2\varphi^2\|\bm{s}\|^2}}{\varphi\|\bm{s}\|+\nu(t)},\ \hat{\theta}(0)\geq0
\end{aligned}
\end{equation}
where $k$, $\sigma_1>0$ and $\sigma_2>0$ are positive design parameters, $\hbar(\zeta)$ is a \emph{BL}-type Nussbaum function, $\hat{\theta}$ is the estimation of $\theta=\textrm{max}\{a_1,\, a_1a_f,\, a_2,\, \bar{y}\}$, and
\begin{equation}\label{core}
\varphi(\bm{\cdot})=\varphi_1(\|\bm{\Phi}\|+\varphi_f+1)+\frac{1}{2}\varphi_2\|\bm{s}\|
\end{equation}
 is the ``core function" (computable and implementable), $\nu(t)$ is a function chosen to satisfy $\nu(t)\geq0$ and $0\leq{\int^t_0}\nu(\tau)d\tau \leq \bar{\nu}< \infty$, $\forall t\geq 0$.
In addition, the following lemma is needed for stability analysis.
\begin{lemma}\label{lemma4}\cite{Wang2010acis}
  Consider the following dynamic system
  \begin{equation}
    \dot{\hat{\gamma}}(t)=-\chi_1\hat{\gamma}(t)+\chi_2\varsigma(t)\notag
  \end{equation}
  where $\chi_1\geq0$, $\chi_2>0$ are constants, and $\varsigma(t)$ is a positive function.
  Then, it holds that $\hat{\gamma}(t)\geq0, \forall t\geq 0$ for any given bounded initial condition $\hat{\gamma}(0)\geq0$.
\end{lemma}

\begin{theorem}\label{theorem1}
  Consider the uncertain nonlinear system (\ref{os}) in the absence of actuator faults.
  Suppose that \emph{Assumptions} \ref{assum3}-\ref{assum2} hold.
  If the controller (\ref{u1}) with the adaptive law (\ref{a1}) is applied, then it holds that
  \begin{enumerate}
    \item all signals in the closed-loop system are GUUB;
    \item the original error $\bm{e}(t)$ asymptotically converges to zero.
  \end{enumerate}
\end{theorem}

\emph{\textbf{Proof:}} Consider the following Lyapunov candidate function
\begin{equation}\label{v1}
  V_1=\frac{1}{2}\bm{s}^T\bm{\alpha} \bm{s}
\end{equation}
{where $\bm{\alpha}\in \mathbb{R}^{m\times m}$ is an unknown symmetric and positive definite matrix under \emph{Assumptions} \ref{assum1}.}
 Taking the time derivative of $V_1$ along (\ref{ds}) yields
\begin{equation}
\begin{aligned}\label{dv1}
\dot{V}_1=\bm{s}^T\bm{\alpha}\big(\bm{g}\bm{u}_a+\bm{\Phi}+ \bm{f}+\bm{d}-\bm{y}^{*(n)}\big) +\frac{1}{2}\bm{s}^T\dot{\bm{\alpha}}\bm{s}.
\end{aligned}
\end{equation}
With \emph{Assumptions} \ref{assum2} and (\ref{L1}), it is straightforward to derive that
\begin{equation}
\begin{aligned}\label{abs}
\|\bm{\alpha}\big(\bm{\Phi}+\bm{f}+\bm{d}- \bm{y}^{*(n)}\big)+\frac{1}{2}\dot{\bm{\alpha}}\bm{s}\|\leq \theta \varphi(\bm{\cdot})
\end{aligned}
\end{equation}
  with $\theta=\textrm{max}\{a_1,\, a_1a_f,\, a_2,\, \bar{y}\}$ and $\varphi(\bm{\cdot})=\varphi_1(\|\bm{\Phi}\|+\varphi_f+1)+\frac{1}{2}\varphi_2\|\bm{s}\|$.
  Then, upon applying \emph{Lemma} \ref{lemma1} and inserting (\ref{u1}), we get
 \begin{equation}
\begin{aligned}\label{dv11}
\dot{V}_1\leq&\, \theta \varphi\|\bm{s}\|+\bm{s}^T\bm{\alpha}\bm{g}\hbar(\zeta) \bigg(k\bm{s}+\frac{{\hat\theta\varphi^2\bm{s}}}{\varphi\|\bm{s}\|+\nu(t)}\bigg)\\
\leq& \ \bm{s}^T\bm{\alpha} \bm{g}\bm{s}\hbar(\zeta)\bigg(k+\frac{{\hat{\theta}\varphi^2}}{\varphi\|\bm{s}\|+\nu(t)}\bigg)+\theta\nu(t)\\
                 &+\theta\frac{{\varphi^2\|\bm{s}\|^2}}{\varphi\|\bm{s}\|+\nu(t)}.
\end{aligned}
\end{equation}
Note that
 \begin{equation}
\label{sgs}
\bm{s}^T\bm{\alpha} \bm{g}\bm{s}=\, \frac{\bm{s}^T\big({\bm{\alpha} \bm{g}+\bm{g}^T\bm{\alpha}}\big)\bm{s}}{2}
+\frac{\bm{s}^T\big({\bm{\alpha} \bm{g}-\bm{g}^T\bm{\alpha}}\big)\bm{s}}{2},
\end{equation}
where ${\bm{\alpha} \bm{g}+\bm{g}^T\bm{\alpha}}$ is a symmetric matrix and ${\bm{\alpha} \bm{g}-\bm{g}^T\bm{\alpha}}$ is a skew-symmetric matrix. Then, for any $\bm{s}\neq0$, it naturally holds that
\begin{equation}\label{beta1}
  \beta_1(t)\triangleq\frac{{\bm{s}}^T\big({\bm{\alpha} \bm{g}+\bm{g}^T\bm{\alpha}}\big){\bm{s}}}{2{\bm{s}}^T{\bm{s}}}\neq0.
\end{equation}
According to \emph{Assumption} \ref{assum1} and \emph{Lemma} \ref{lemma5}, there must exist two constants $\underline{\lambda}$ and $\bar{\lambda}$ such that
\begin{equation}\label{ineqbeta}
  \underline{\lambda}\leq \lambda_m(t) \leq 2\beta_1(t)\leq \lambda_M(t) \leq \bar{\lambda}.
\end{equation}
Thus, it holds that
\begin{equation}\label{betas}
{{\bm{s}}^T\big({\bm{\alpha} \bm{g}+\bm{g}^T\bm{\alpha}}\big){\bm{s}}}={2}\beta_1(t)\|\bm{s}\|^2.
\end{equation}
 In addition, the condition $\lambda_m\lambda_M>0$ in \emph{Assumption} \ref{assum1} implies that the sign of $\beta_1(t)$ is strictly positive or strictly negative, but unknown.

Using (\ref{sgs}), (\ref{betas}), and the fact that $\bm{s}^T\big({\bm{\alpha} \bm{g}-\bm{g}^T\bm{\alpha}}\big)\bm{s}=0$, we obtain
 \begin{equation}
\begin{aligned}\label{sgs1}
\bm{s}^T\bm{\alpha} \bm{g}\bm{s}=\beta_1(t)\|\bm{s}\|^2.
\end{aligned}
\end{equation}
Subsequently, using \emph{Lemma} \ref{lemma4}, and inserting (\ref{sgs1}) into (\ref{dv11}) yields
 \begin{equation}
\begin{aligned}\label{dv12}
\dot{V}_1
\leq &\ \beta_1(t)\hbar(\zeta)\bigg(k\|\bm{s}\|^2+\frac{{\hat{\theta}\varphi^2}\|\bm{s}\|^2}{\varphi\|\bm{s}\|+\nu(t)}\bigg)+\theta\frac{{\varphi^2\|\bm{s}\|^2}}{\varphi\|\bm{s}\|+\nu(t)}+\theta\nu(t) \\
\leq &-k\|\bm{s}\|^2+\big(\beta_1(t)\hbar(\zeta)+1\big)\bigg(k\|\bm{s}\|^2+\frac{{\hat{\theta}\varphi^2}\|\bm{s}\|^2}{\varphi\|\bm{s}\|+\nu(t)}\bigg)\\
&+(\theta-\hat\theta)\frac{{\varphi^2\|\bm{s}\|^2}}{\varphi\|\bm{s}\|+\nu(t)}+\theta \nu(t).
\end{aligned}
\end{equation}
 Now we introduce a virtual parameter estimation error of the form $\tilde{\theta}=\theta-\hat\theta$,
then blend such error into the second part of the complete Lyapunov function candidate such that
\begin{equation}
\begin{aligned}\label{v2}
V_2=V_1+\frac{1}{2\sigma_2}{\tilde{\theta}}^2.
\end{aligned}
\end{equation}
Differentiating (\ref{v2}) and using (\ref{dv12}), yields
\begin{equation}
\begin{aligned}\label{dv2}
\dot{V}_2\leq &-k\|\bm{s}\|^2+\big(\beta_1(t)\hbar(\zeta)+1\big)\bigg(k\|\bm{s}\|^2+\frac{{\hat{\theta}\varphi^2}\|\bm{s}\|^2}{\varphi\|\bm{s}\|+\nu(t)}\bigg)\\
&+\tilde{\theta}\bigg(\frac{{\varphi^2\|\bm{s}\|^2}}{\varphi\|\bm{s}\|+\nu(t)}-\frac{\dot{\hat{\theta}}}{\sigma_2}\bigg)+\theta \nu(t).
\end{aligned}
\end{equation}
By inserting (\ref{a1}) into (\ref{dv2}), we can further get
 \begin{equation}
\begin{aligned}\label{dv22}
\dot{V}_2\leq -k\|\bm{s}\|^2+\frac{1}{\sigma_1}\big(\beta_1(t)\hbar(\zeta)+1\big)\dot{\zeta}+\theta \nu(t).
\end{aligned}
\end{equation}
Integrating both sides of (\ref{dv22}) yields that
 \begin{equation}
\begin{aligned}\label{idv22}
{V}_2(t)+ k{\int_{0}^{t}}\|\bm{s}(\tau)\|^2d\tau \leq \frac{1}{\sigma_1}{\int_{0}^{t}}\big(\beta_1(\tau)\hbar(\zeta(\tau))+1\big)\dot{\zeta}(\tau)d\tau+\Delta,
\end{aligned}
\end{equation}
where $\Delta={V}_2(0)+\theta \bar{\nu}$.
Therefore, from \emph{Lemma} \ref{lemma0}, and the definition of $V_2$ in (\ref{v2}) along with (\ref{dv22}) and (\ref{idv22}), we establish that $V_2\in {\mathcal{L}_\infty}$, $\zeta\in {\mathcal{L}_\infty}$, $\hat{\theta}\in {\mathcal{L}_\infty}$, and $\bm{s}\in {\mathcal{L}_\infty}\cap{\mathcal{L}_2}$.
From \emph{Lemma} \ref{lemma2} and (\ref{z}), $\bm{s}\in {\mathcal{L}_\infty}$ implies that $\bm{e}\in {\mathcal{L}_\infty}$ and $\bm{e}_i\in {\mathcal{L}_\infty}$ ($i=2,\cdots,n$).
According to \emph{Assumption} \ref{assum3} and the definition of $\bm{e}$ and $\bm{s}$, it follows that $\bm{x}_i \in {\mathcal{L}_\infty}$ $(i=1,\cdots,n)$, then $\varphi(\bm{\cdot}) \in{\mathcal{L}_\infty}$ by \emph{Assumption}  \ref{assum1} and  \ref{assum2}.
Then $\bm{u}_a\in {\mathcal{L}_\infty}$ and $\dot{\hat{\theta}}\in{\mathcal{L}_\infty}$ from (\ref{u1}) and (\ref{a1}).
Finally, from (\ref{ds}), one can conclude that $\dot{\bm{s}}\in{\mathcal{L}_\infty}$.
Therefore, the GUUB for all the signals in the closed-loop adaptive system is guaranteed.
In addition, since $\bm{s}\in {\mathcal{L}_\infty}\cap{\mathcal{L}_2}$ and $\dot{\bm{s}}\in{\mathcal{L}_\infty}$, by applying Barbalat's lemma, we have $\textrm{lim}_{t\rightarrow\infty}\bm{s}(t)=\bm{0}_m$.
Therefore, from \emph{Lemma} \ref{lemma2}, the original error $\bm{e}(t)$ and its derivative up to $n$th order converge asymptotically to zero as $t\rightarrow \infty$ with the same decreasing rate as that of $\bm{s}(t)$.
\QEDB
\begin{remark}\label{Remark7}
  {Different from the work in \cite{Lee2016scl} that directly utilizes the auxiliary matrix $\bm{\alpha}$ in the control design, here we just use it to construct the Lyapunov function (\ref{v1}). In order to cope with the resultant extra term $\bm{s}^T\dot{\bm{\alpha}}\bm{s}$ in the stability analysis, the readily computable term $\|\bm{s}\|$ is incorporated into the robust unit (\ref{u1}) and the adaptive unit (\ref{a1}) of the control scheme (whether $\varphi_1$ and $\varphi_2$ can be extracted or not), which thus enables the proposed control scheme to exhibit stronger robustness against the system structure under less conservative  controllability conditions. It is also noted that when $\bm{\alpha} = \bm{I}_m$, the control scheme (\ref{u1})-(\ref{a1}) reduces to the control scheme in \cite{Song-Huang2016tac} since the term $\|\bm{s}\|$ will not appear in $\varphi(\bm{\cdot})$ any more.}
\end{remark}

\begin{remark}\label{Remark8}
{It should be noted that the results for uncertain nonlinear systems with unknown control direction and time-varying control gains in \cite{Ge2003tac,Ge2004smc,Bechlioulis2009auto,Jiang2006auto} are only valid on a finite time interval $[0, t_f)$. However, the developed control scheme is able to achieve the same result for all $t\geq 0$ by using \emph{BL}-type Nussbaum functions.}
\end{remark}

\section{Control With Actuator Faults}
\subsection{Control Design With Actuator Faults}
As unanticipated actuator faults may occur for ``long-term" operation,  the actual control input $\bm{u}_a$ and the designed control signal $\bm u=[u_{1},\ldots,u_{m}]^T \in \mathbb{R}^m$ in this scenario are not identical anymore.
Instead, the abnormal actuator input-output model is described as
 \begin{equation}
 \label{af}
 \bm{u}_{a}=\bm{\rho}(t)\bm{u}+\bm{\varepsilon}(t)
\end{equation}
where $\bm{\rho}=\textmd{diag}\{\rho_1,\ldots,\rho_m\} \in \mathbb{R}^{m \times m}$ is the coefficient matrix of unknown multiplicative actuator faults and can be piecewise continuous that reflects the effectiveness of actuators; $\bm{\varepsilon}=[\varepsilon_1,\ldots,\varepsilon_m]^T \in \mathbb{R}^{m}$ is the uncontrollable portion of actuator input vector, which is assumed to be unknown but bounded, such that $\|\bm{\varepsilon}\|<\bar{\varepsilon}<\infty$, where $\bar{\varepsilon}$ being a positive constant.
In this subsection, we consider the PLOE case, that is the actuators are always functional such that $\bm{u}_a$ can be influenced by the control input $\bm{u}$ all the time (i.e., $\rho_j \in (0,1], j=1,\ldots,m$).

To obtain an expression of the system with actuator faults, Eqs. (\ref{os}) and (\ref{af}) are rearranged by
\begin{equation}
\begin{cases}
\label{afs}
\dot{\bm{x}}_{i}=\bm{x}_{i+1},\quad i=1,2,\ldots,n-1, \\
\dot{\bm{x}}_{n}=\bm{f}(\bar{\bm{x}})+\bm{g}(\bar{\bm{x}},t)\bm{\varepsilon}(t)+{\bm{d}}(\bar{\bm{x}},t)+\bm{g}(\bar{\bm x},t)\bm{\rho}(t)\bm{u},\\
\bm{y}=\bm{x}_1.
\end{cases}
\end{equation}

To proceed, the following assumptions need to be extended on the basis of \emph{Assumptions} \ref{assum1} and \ref{assum2}, respectively.

\begin{assumption}\label{assum4}
  {For uncertain system (\ref{afs}) with multiplicative actuator faults, there exists an unknown positive and symmetric matrix $\bm{\alpha}(\bar{\bm{x}}_{n-1},t) \in \mathbb{R}^{m \times m}$ which is differentiable with respect to $t$ and ${\bm{x}}_j$ ($j=1,\ldots,n-1$), such that ${{\bm{\alpha} \bm{g\rho}+\bm{\rho}\bm{g}^T\bm{\alpha}}}$ is either positive or negative definite, i.e., $\lambda^*_m\lambda^*_M>0$ with $\lambda^*_m(t):=\textrm{min}\{\textrm{eig}({{\bm{\alpha} \bm{g\rho}+\bm{\rho}\bm{g}^T\bm{\alpha}}})\}$ and $\lambda^*_M(t):=\textrm{max}\{\textrm{eig}({{\bm{\alpha} \bm{g\rho}+\bm{\rho}\bm{g}^T\bm{\alpha}}})\}$.}
\end{assumption}
\begin{assumption}\label{assum5}
  There exist an unknown positive constant $a^*_f$ and known nonnegative scalar function $\varphi^*_f (\bar{\bm{x}})$, such that
  \begin{equation}\label{afphi*}
    \|\bm{f}(\bar{\bm{x}})+\bm{g}(\bar{\bm{x}},t)\bm{\varepsilon}(t)+\bm{d}(\bar{\bm{x}},t)\|\leq a^*_f \varphi^*_f(\bar{\bm{x}})
  \end{equation}
  where $\varphi^*_f(\bar{\bm{x}})$ is radially unbounded. {In addition, it is also assumed that the condition (\ref{aphi}) in \emph{Assumption} \ref{assum2} still holds.}
\end{assumption}
\begin{remark}\label{Remark9}
  {\emph{Assumption} \ref{assum4} is the controllability condition of MIMO systems with multiplicative faults.
  In \cite{Jin2019tcyb}, ${\bm{g\rho}+\bm{\rho}\bm{g}^T}$ is assumed to be either negative definite or positive definite directly.
  However, as noted in \cite{Zhang-Yang2020auto}, such assumption would likely be invalidated in some scenarios.
  In \cite{Huang-Song2021tac,Zhang-Yang2020auto}, similar conditions are considered, yet $\bm{\alpha}$ is required to be diagonal matrix in \cite{Huang-Song2021tac}, and $\bm{\alpha}=\bm{\rho}$ is established in \cite{Zhang-Yang2020auto}, which thus cannot handle intermittent actuator faults.
  Clearly, those assumptions in \cite{Jin2019tcyb,Huang-Song2021tac,Zhang-Yang2020auto} can be essentially regarded as some special cases of \emph{Assumption} \ref{assum4}. }
\end{remark}

Consider now the generalized error dynamics (\ref{ds}) in the presence of intermittent actuator faults with $\bm{u}_{a}$ given by (\ref{af}), which yields
\begin{equation}
\label{dsf}
\dot{\bm{s}}=\bm{\Phi}+\bm{g\rho u}+\bm{f}+\bm{g\varepsilon }+\bm{d}- \bm{y}^{*(n)}.
\end{equation}
Next, design the fault-tolerant control scheme as
\begin{equation}
\begin{aligned}\label{fu2}
\bm{u}&=\hbar(\zeta)\bm{\eta}, \\
\bm{\eta}&=k\bm{s}+\frac{{\hat\theta\varphi^2\bm{s}}}{\varphi\|\bm{s}\|+\nu(t)}
\end{aligned}
\end{equation}
with
\begin{equation}
\begin{aligned}\label{fa2}
\dot{\zeta}&=\sigma_1\bm{s}^T\bm{\eta},\\
\dot{\hat{\theta}}&=\frac{{\sigma_2\varphi^2\|\bm{s}\|^2}}{\varphi\|\bm{s}\|+\nu(t)},\ \hat{\theta}(0)\geq0
\end{aligned}
\end{equation}
where $\hat{\theta}$ is the estimation of $\theta=\textrm{max}\{a_1,\, a_1a^*_f,\, a_2,\, \bar{y}\}$ and $\varphi(\bm{\cdot})=\varphi_1(\|\bm{\Phi}\|+\varphi^*_f+1)+\frac{1}{2}\varphi_2\|\bm{s}\|$.
\begin{theorem}\label{theorem2}
  Consider the uncertain MIMO nonlinear system (\ref{afs}) in the presence of actuator faults. Suppose that \emph{Assumptions} \ref{assum3}, \ref{assum4} and \ref{assum5} hold.
  If the control algorithm (\ref{fu2}) and (\ref{fa2}) are applied, then similar results as \emph{Theorem} \ref{theorem1} hold.
\end{theorem}

\emph{\textbf{Proof:}} We first choose the Lyapunov candidate function as $V_2=\frac{1}{2}\bm{s}^T\bm{\alpha} \bm{s}+\frac{1}{2\sigma_2}{\tilde{\theta}}^2$, where $\tilde{\theta}$ is the virtual parameter estimation error defined as $\tilde{\theta}=\theta-\hat\theta$.
Then, taking the derivative of $V_2$ along (\ref{dsf}) as
\begin{equation}
\begin{aligned}\label{dv3}
\dot{V}_2=\bm{s}^T\bm{\Theta} +\bm{s}^T\bm{\alpha} \bm{g}\bm{\rho} \bm{u}+\frac{1}{\sigma_2}\tilde{\theta}\dot{\tilde{\theta}}
\end{aligned}
\end{equation}
where $\bm{\Theta} =\bm{\alpha}\big(\bm{\Phi}+(\bm{f}+\bm{g\varepsilon} +\bm{d})-\bm{y}^{*(n)}\big)+\frac{1}{2}\bm{s}^T\dot{\bm{\alpha}}\bm{s}$.
Following \emph{Assumptions} \ref{assum3} and \ref{assum5} and using \emph{Lemma} \ref{lemma1}, we have
\begin{equation}
\begin{aligned}\label{Theta}
\bm{s}^T\bm{\Theta} \leq \|\bm{s}\|\theta \varphi \leq \theta \nu(t)+\theta\frac{{\varphi^2\|\bm{s}\|^2}}{\varphi\|\bm{s}\|+\nu(t)}.
\end{aligned}
\end{equation}
Substituting (\ref{fu2}) and (\ref{Theta}) into (\ref{dv3}) yields
 \begin{equation}
\begin{aligned}\label{dv30}
\dot{V}_2\leq& \bm{s}^T\bm{\alpha} \bm{g}\bm{\rho}\bm{s}\hbar(\zeta)\bigg(k+\frac{{\hat{\theta}\varphi^2}}{\varphi\|\bm{s}\|+\nu(t)}\bigg)+\theta \nu(t)-\frac{1}{\sigma_2}\tilde{\theta}\dot{\hat{\theta}}.
\end{aligned}
\end{equation}
Note that
\begin{equation}
\begin{aligned}\label{sgrs}
\bm{s}^T\bm{\alpha} \bm{g\rho}\bm{s}=\, \frac{\bm{s}^T\big({\bm{\alpha} \bm{g\rho}+\bm{\rho g}^T\bm{\alpha}}\big)\bm{s}}{2}
+\frac{\bm{s}^T\big({\bm{\alpha} \bm{g\rho}-\bm{\rho g}^T\bm{\alpha}}\big)\bm{s}}{2},
\end{aligned}
\end{equation}
with ${\bm{\alpha}\bm{g\rho}+\bm{\rho}\bm{g}^T\bm{\alpha}}$ being symmetric and ${\bm{\alpha} \bm{g\rho}-\bm{\rho}\bm{g}^T\bm{\alpha}}$ being skew-symmetric matrix. Then, from \emph{Assumption} \ref{assum4} and \emph{Lemma} \ref{lemma5}, it holds that
 \begin{equation}
\begin{aligned}\label{sGs1}
\bm{s}^T\bm{\alpha} \bm{g\rho}\bm{s}=\beta_2(t)\|\bm{s}\|^2,
\end{aligned}
\end{equation}
where $\beta_2(t)\neq0$ for any $t\in[0,\infty)$, and the sign of $\beta_2(t)$ is unknown.

Inserting (\ref{sGs1}) and (\ref{fa2}) into (\ref{dv30}), and using \emph{Lemma} \ref{lemma4}, $\dot{V}_2$ can be further bounded as
 \begin{equation}
\begin{aligned}\label{dv31}
\dot{V}_2 \leq -k\|\bm{s}\|^2+\frac{1}{\sigma_1}\big(\beta_2(t)\hbar(\zeta)+1\big)\dot{\zeta}+\theta \nu(t).
\end{aligned}
\end{equation}

Integrating both sides of (\ref{dv31}) yields that
 \begin{equation}
\begin{aligned}\label{idv42}
{V}_2(t)+ k{\int_{0}^{t}}\|\bm{s}(\tau)\|^2d\tau \leq \frac{1}{\sigma_1}{\int_{0}^{t}}\big(\beta_2(\tau)\hbar(\zeta(\tau))+1\big)\dot{\zeta}(\tau)d\tau+\Delta,
\end{aligned}
\end{equation}
where $\Delta={V}_2(0)+\theta \bar{\nu}$.
Finally, by following the similar analysis as that used in the proof of \emph{Theorem} \ref{theorem1}, it is readily shown that all the signals in the closed-loop system are GUUB, and the original error $\bm{e}(t)$ and its derivative up to $n$th order converge asymptotically to zero as $t\rightarrow \infty$ with the same decreasing rate as that of $\bm{s}(t)$.
\QEDB
\begin{remark}\label{Remark10}
  {In the presence of actuator faults, the relaxation of the controllability condition could be more involved due to the coupling between the fault coefficient matrix $\bm{\rho}$ and the input gain $\bm{g}$. In such case, the existence of the auxiliary matrix $\bm{\alpha}$ should be more urgent since the considered actuator faults are undetectable and intermittent (i.e., $\bm{\rho}$ is unknown and discontinuous). However, for some special practical systems, some suitable choices of $\bm{\alpha}$ can be always established, which can be seen in the next section with the application of the developed scheme to robotic systems.}
\end{remark}

\begin{remark}\label{Remark11}
{It should be stressed that for narrative concision and  comprehension, we consider the system model in normal form \cite{Bechlioulis2017tac,Kanakis2020tac}, however, the method can be extended straightforwardly to more general systems, like nonaffine systems \cite{Song-Huang2016tac}, strict feedback systems \cite{Zhao-Song2019auto} and pure feedback systems \cite{Huang-Song2020auto}.
For example, extension to the nonaffine systems with the canonical form $\dot {\bm{x}}=\bm{f}(\bm{x},\bm{u})$ is immediate by using the mean value theorem.
Nevertheless, for systems in triangular form, such extension could be more involved since the backstepping design technique could be used for control design.
In such a case, the controllability condition in \emph{Assumption} \ref{assum1} on the auxiliary term $\bm{\alpha}$ need to be updated correspondingly due to the presence of the virtual controllers, which represents an interesting topic for further research.}
 \end{remark}

\subsection{Numerical Example}
\begin{figure}[!htp]
  \centering
  \includegraphics[width=7.5cm]{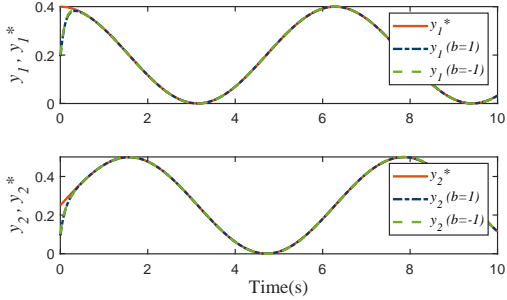}
  \caption{Position tracking process.}
  \label{Nposition}
\end{figure}

\begin{figure}[!htp]
  \centering
  \includegraphics[width=7.5cm]{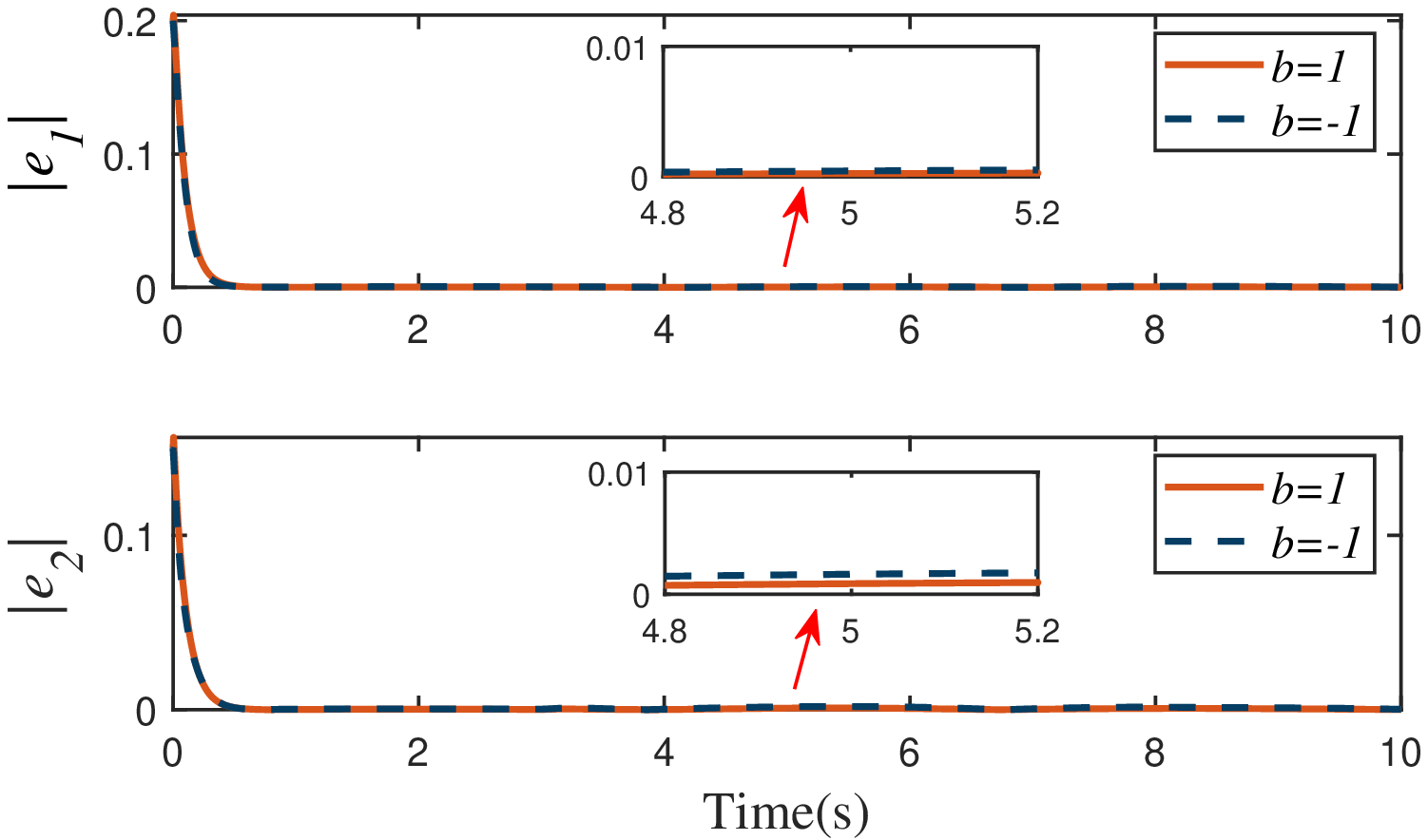}
  \caption{Tracking errors.}
  \label{Nerror}
\end{figure}

\begin{figure}[!htp]
  \centering
  \includegraphics[width=7.5cm]{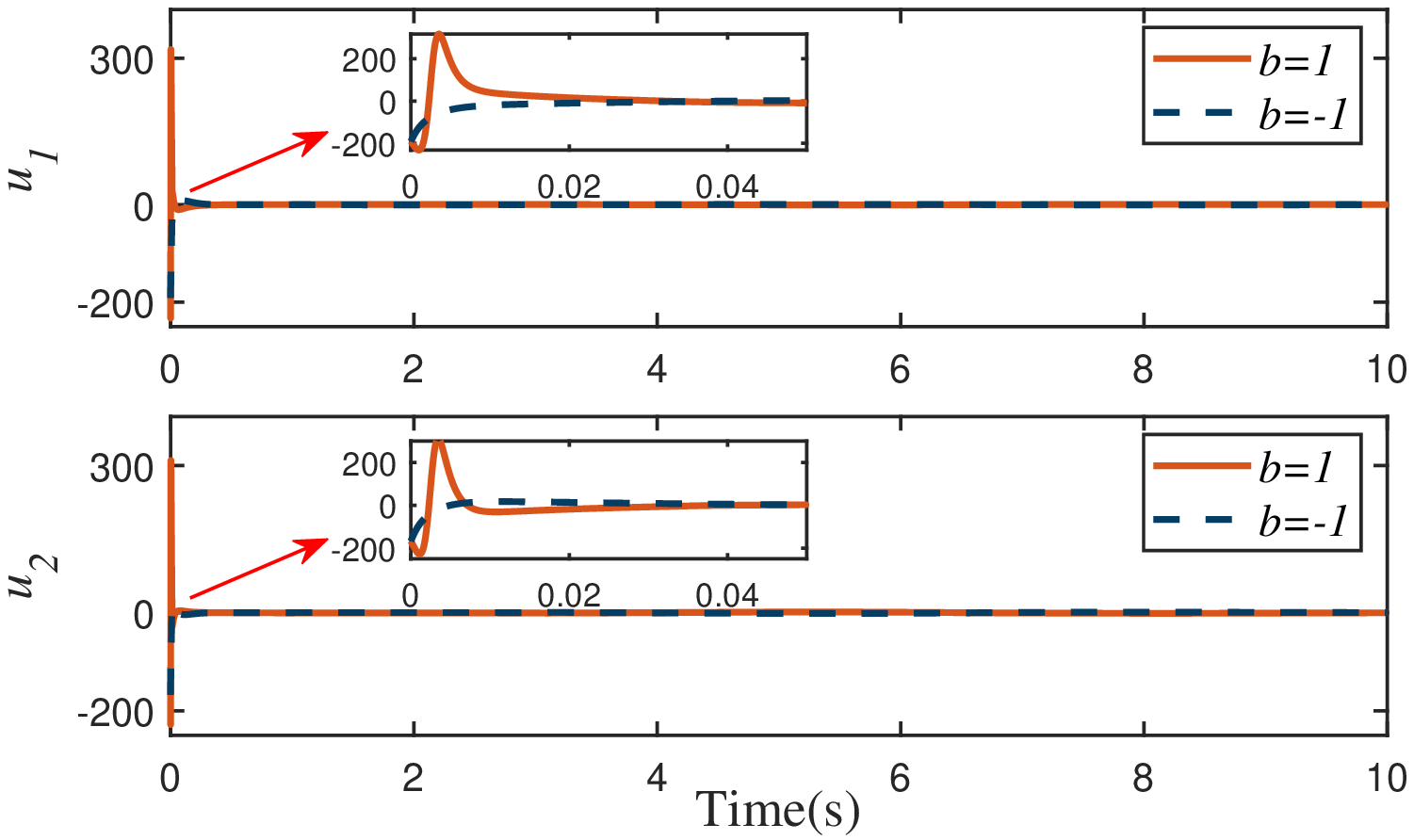}
  \caption{Control input signals.}
  \label{Ntau}
\end{figure}

\begin{figure}[!htp]
  \centering
  \includegraphics[width=7.5cm]{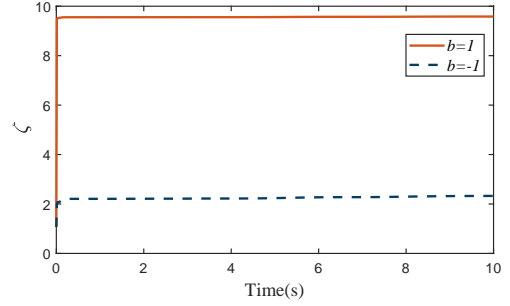}
  \caption{Adapting parameter $\zeta$.}
  \label{Nzeta}
\end{figure}

To test the effectiveness of the proposed control scheme, we consider the system (\ref{afs}) as
\begin{equation}\label{afs1}
\begin{cases}
\dot{\bm{x}}_{1}=\bm{x}_{2} \\
\dot{\bm{x}}_{2}=\bm{f}(\bar{\bm{x}})+b\bm{g}(\bar{\bm{x}},t)\bm{\rho }(t)\bm{u}_a+\bm{d}(\bar{\bm{x}},t)\\
\bm{y}=\bm{x}_1.
\end{cases}
\end{equation}
with
$\bm{f}(\bar{\bm{x}})=[0.5x_{21}\sin(x_{11});0.6x_{22}\tanh(x_{21})]$, $\bm{g}(\bar{\bm{x}},t)=[2+0.1\cos(t),2+0.1\cos(x_{11}x_{21});2+0.1\sin(x_{12}x_{22}),3+0.1\sin(t)]$ and $\bm{d}(\bm{\bar{x}}_2,t) = [0.05x_{11}x_{21}\sin(t);0.05x_{12}x_{22}\sin(t)]$, where $\bar{\bm{x}}=[\bm{x}_1^T,\bm{x}_2^T]^T$ and $b=\{-1, 1\}$.
The jump-type PLOE actuator faults are set as
    \begin{equation}
\begin{aligned}
       &\bm{\rho}(t)= \begin{cases} \textmd{diag}\{0.9+0.1\sin(t),1-0.2\tanh(t)\},&t\in(0,3]\\
    \textmd{diag}\{0.8+0.2\sin(t),0.2\},&t\in(3,\infty)\end{cases}\\
       & \bm{\varepsilon}(t)=[0.02\tanh(2t);0.02\cos(3t)]
  \end{aligned}
  \end{equation}
In such case, the controllability conditions given in \cite{Jin2019tcyb,Zhang-Yang2020auto} are not satisfied since ${ \bm{g\rho}+\bm{\rho}\bm{g}^T}$ is neither uniformly positive nor uniformly negative definite for all $t\in[0,\infty)$ and $\bm{\rho}$ is not differentiable at the time instant $t = 3s$. However, \emph{Assumption} \ref{assum4} can be met with $\bm{\alpha}(\bm{x}_1,t)$ given as
  \begin{equation}\label{exa2}
  \bm{\alpha}({\bm{x}}_{1},t)=\begin{bmatrix}
    1+0.1\sin(t) & 0.1\sin(x_{11}) \\
    0.1\sin(x_{11}) & 0.4+0.1\cos(t)
  \end{bmatrix},\  t\in[0,\infty).
  \end{equation}
The design parameters are chosen as $k=100$, $\sigma_1=1$, $\sigma_2=0.1$, $\lambda_1=10$. The integrable function is set as $\nu(t)=0.5e^{-0.5t}$. The ``core" function $\varphi^*_f(\bm{\cdot})$ is chosen as $\varphi^*_f(\bm{\cdot})=\|\bm{x}_1\|\|\bm{x}_2\|+\|\bm{x}_2\|+1$.
The \emph{BL}-type Nussbaum function is chosen as $\hbar(\zeta)=\exp(0.07\zeta^2)\cos(0.1\pi\zeta)$.
The desired signal is given as $\bm{y}^*=[0.2+0.2\cos(t);0.25+0.25\sin(t)]$, and the initial conditions are: $\bm{x}_1(0)=[0.2;0.1]$,  $\bm{x}_2(0)=[0;0]$, $\dot{\zeta}(0)=1$, $\hat{\theta}(0)=0$. {In addition, two different control directions (i.e., $b=1$ and $b=-1$) are considered with the same set of design parameters, which is essentially different from the current works \cite{Ge2004smc,Liu2006tac,Bechlioulis2009auto,Shi2014tfs,Jiang2006auto} that test their results under single control direction. The simulation results are shown in Figs. \ref{Nposition}-\ref{Nzeta}. It can be seen from Figs. \ref{Nposition} and \ref{Nerror} that the system output $\bm{y}$ well tracks the desired trajectory $\bm{y}^*$ and the tracking error $\bm{e}$ converges to zero. The boundedness of the control input signal $\bm{u}$ and the adapting parameter is illustrated in Figs. \ref{Ntau} and \ref{Nzeta}. Thus, although the system (\ref{afs1}) does not meet the controllability conditions given in \cite{Jin2019tcyb,Zhang-Yang2020auto} and the control direction is unknown, the proposed control method can still automatically accommodate the jump-type PLOE actuator faults and achieve asymptotic tracking, moreover, it is effective under both control directions.}


\section{Application to Robotic Systems}
\subsection{Control Design for Robotic Systems}
To examine the applicability and feasibility of the proposed method, we consider an $m$ degree-of-freedom (DOF) rigid-link robotic manipulator with the following dynamics:
\begin{equation}
\label{dy}
{\textsl{M}}(\bm{q})\ddot {\bm{q}}+{\textsl{C}}(\bm{q},\dot {\bm{q}})\dot { \bm{q}}+{\textsl{G}}(\bm{q})+{\bm{\tau}_d(\dot{\bm{q}},t)} = \bm{u}_a,
\end{equation}
where ${\bm{q}}\in \mathbb{R}^m$ denote the joint position, ${\textsl{M}}({\bm{q}})\in \mathbb{R}^{m\times m}$ denotes the inertia matrix which is symmetric and positive definite, ${\textsl{C}}({\bm{q}},\dot {\bm{q}})\in \mathbb{R}^{m\times m}$ represents the centripetal-coriolis matrix, ${\textsl{G}}({\bm{q}})\in \mathbb{R}^m$ is the vector of gravitational force; ${\bm{u}}_a \in {R}^m$ is the actual joint control torque, and ${\bm{\tau}_d(\dot{{\bm{q}}},t)} \in \mathbb{R}^m$ is the external disturbance input. By taking ${\bm{x}}_1={\bm{q}}$ and ${\bm{x}}_2=\dot{{\bm{q}}}$, {and considering  the abnormal actuator input-output
model as (\ref{af}),} the dynamics (\ref{dy}) can be transformed into the normal form
 \begin{equation}
 \label{tdy}
  \begin{cases} \dot{\bm{x}}_1=\bm{x}_2\\
    \dot{\bm{x}}_2=\bm{f}+\bm{d}+\bm{g\varepsilon} +\bm{g\rho u } \end{cases}
  \end{equation}
 where $\bm{g}={\textsl{M}}^{-1}$ and $\bm{f}+\bm{d}+\bm{g\varepsilon} =\bm{g}(-{\textsl{C}}\dot{\bm{q}}-{\textsl{G}}-{\bm{\tau}_d}+\bm{\varepsilon})$.
 The subsequent is based on the assumption that ${\bm{x}}_1$ and ${\bm{x}}_2$ are measurable and ${\textsl{M}}({\bm{\cdot}})$, ${\textsl{C}}({\bm{\cdot}})$, ${\textsl{G}}({\bm{\cdot}})$ and ${\bm{\tau}_d(\bm{\cdot})}$ are unknown.

 Let ${\bm{y}}={\bm{x}}_1$ and ${\bm{y}}^*={\bm{q}}_d$ be the desired trajectory, and the tracking error is defined as
${\bm{e}}={\bm{y}}-{\bm{y}}^*={\bm{q}}-{\bm{q}}_d$.
 Then, we introduce the filtered variable as ${\bm{s}}(t)=\lambda_1{\bm{e}}+\dot{\bm{e}}$.

\emph{\textbf{Corollary} 1:} Suppose that \emph{Assumptions} \ref{assum3},  \ref{assum4} and \ref{assum5} hold, if the control algorithm as (\ref{fu2}) and (\ref{fa2}) are applied,
then the results in \emph{Theorem} 2 still hold for robotic systems (\ref{tdy}).

\emph{\textbf{Proof:}} The proof procedures are similar to those of \emph{Theorem} \ref{theorem2} and are omitted for brevity.
\QEDB
\begin{remark}\label{Remark12}
 {Since the control direction of the robotic system (\ref{dy}) is known, the controllability condition in \emph{Assumptions} \ref{assum4} can be modified as ``{$
    0<\underline{\lambda} \leq\textrm{min}\{\textrm{sv}({{\bm{\alpha} \bm{g\rho}+\bm{\rho}\bm{g}^T\bm{\alpha}}})\}$ where $\underline{\lambda}$ is an unknown constant}''. In such case, it can be easily verified that the proposed control methodology is further simplified to:
    \begin{equation}
    \begin{aligned}\label{ru2}
    \bm{u}&=-k\bm{s}-\frac{{\hat\theta{\varphi}^2\bm{s}}}{{\varphi}\|\bm{s}\|+\nu(t)},\\
    \dot{\hat{\theta}}&=\frac{{\sigma_2{\varphi}^2\|\bm{s}\|^2}}{{\varphi}\|\bm{s}\|+\nu(t)},\ \hat{\theta}(0)\geq0.
    \end{aligned}
    \end{equation}}
\end{remark}
\begin{remark}\label{Remark13}
{Continuing the reasoning of \emph{Remark} \ref{Remark12}, it is interesting to note that for the typical robotic systems, the existence of the auxiliary matrix $\bm{\alpha}$ is diverse and three respective cases can be discussed as follows:
\begin{itemize}
  \item[\emph{i}:] Choose $\bm{\alpha} = \bm{I}_m$.  In this case, the considered controllability condition is equivalent to the traditional one  imposed in \cite{Jin2019tcyb};
  \item[\emph{ii}:] Choose $\bm{\alpha} = \bm{\rho}$. In this case, the considered controllability condition can be boiled down to the one  in \cite{Zhang-Yang2020auto}, which, however, requires $\bm{\rho}$ to be differentiable thus cannot cope with intermittent actuator faults;
  \item[\emph{iii}:] Choose $\bm{\alpha} = \textsl{M}$. In such case, the constructed controllability condition is the same as that in \cite{Cao2020ijc}, which is able to deal with intermittent actuator faults.
\end{itemize}
Therefore, the controllability condition imposed in \cite{Jin2019tcyb, Zhang-Yang2020auto,Cao2020ijc} for robotic systems are essentially some special cases of ours. Clearly, the choice of $\bm{\alpha}$ in \emph{Case 3} is more powerful, which adeptly takes advantage of the inherent properties of the inertia matrix $\textsl{M}$, that is, $\|\textsl{M}\| \leq\lambda^*_M$ and $\|\dot{\textsl{M}}\| \leq k_M\|\dot{{\bm{q}}}\|$, which is exactly inconsistent with the condition (\ref{aphi}) in \emph{Assumption} \ref{assum2}.}
\end{remark}

\subsection{Case Study}
In this subsection, we verify the performance of the proposed control scheme using a 3-DOF rigid-link robotic manipulator system, whose dynamics model borrowed from \cite{Xin2007trob} (see \cite{Xin2007trob} for detail expression and definition), and the parameters for the model are listed in Table. \ref{table}.
\begin{table}[ht]
\captionsetup{font={small}}
\caption{Parameters of 3-DOF rigid-link robotic manipulator.}
\centering
\begin{tabular}{c c c c}
\hline\hline
Link $i$    &{Link 1} &{Link 2} &{Link 3}\\[0.5ex]
\hline
{$m_i$ (kg) }       &0.5       &0.5    &0.5    \\[0.5ex]
 {$l_i$ (m)}       &1.0       &1.0     &1.0   \\[0.5ex]
{$l_{ci}$ (m)}       &0.5       &0.5     &0.5   \\[0.5ex]
{$I_i$ (kg$\cdot$$\textmd{m}^2$) }       &1.5       &1.0     &0.5   \\[0.5ex]
\hline
\end{tabular}
\label{table}
\end{table}

The external disturbance input is given as
\begin{equation}
  \bm{\tau}_d(\dot{\bm{q}},t) = [0.02\dot{q}_1\sin(t);0.02\dot{q}_2\cos(t);0.02\dot{q}_3\sin(t)].
\end{equation}
The three control channels of the system are subject to both additive and actuation effectiveness faults, which are determined by
  \begin{equation}
  \begin{aligned}
      &\bm{\rho}(t)=\begin{cases}\textmd{diag}\big\{1-0.2\tanh(t),0.9+0.1\sin(t),\\
      \qquad\ 0.9+0.1\cos(t)\big\},\ \ \ \ \ \, t\in(0,5] \\
      \textmd{diag}\big\{0.8+0.05\sin(t),0.2+0.05\cos(t),\\
      \qquad \  0.2-0.05\tanh(t)\big\},\ \ t\in(5,\infty)\end{cases}\\
     & \bm{\varepsilon}(t)= [0.01\tanh(2t);0.01\cos(t);0.01\sin(3t)].
  \end{aligned}
  \end{equation}
In this case, the matrix ${\bm{g\rho}+\bm{\rho}\bm{g}^T}$ is not positive or negative definite and $\bm{\rho}$ is not differentiable at the time instant $t = 5s$. However, if we choose $\bm{\alpha}(\bm{\cdot})$ as
\begin{equation} \label{Alpha}
\setlength{\arraycolsep}{1pt}
  \begin{array}{lc}
\bm{\alpha}(\bm{q},t)\\
=\begin{bmatrix} 0.6+0.05\sin(t)&0.01\sin(q_{1})&0\\ 0.01\sin(q_{1})&0.2+0.05\cos(t)&0.01\sin(q_{2})\\
  0&0.01\sin(q_{2})&0.2-0.05\tanh(t)\end{bmatrix},
  \end{array}
\end{equation}
then the $\emph{Assumption}$ \ref{assum4} is met. The desired trajectory is set as
\begin{equation}
  \bm{q}_d=[0.5+0.5\sin(t);0.5-0.5\cos(t);0.5-0.5\sin(t)].
\end{equation}

\begin{figure}[!htp]
  \centering
  \includegraphics[width=7.5cm]{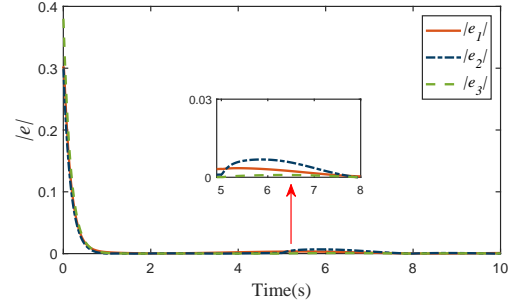}
  \caption{Tracking errors.}
  \label{error3}
\end{figure}

\begin{figure}[!htp]
  \centering
  \includegraphics[width=7.5cm]{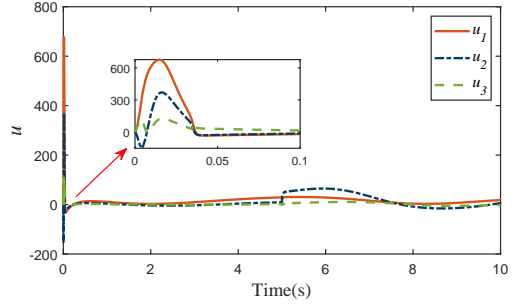}
  \caption{Control input signals.}
  \label{tau3}
\end{figure}

The design procedure for our proposed method is given in \emph{Section} V-A.
The design parameters are chosen as $k=100$, $\sigma_1=0.1$, $\lambda_1=20$.
The integrable function is set as $\nu(t)=0.5e^{-0.5t}$.
The ``core" function $\varphi(\bm{\cdot})$ is chosen as $\varphi(\bm{\cdot})=\|\Phi\|+\varphi^*_f(\bm{\cdot})+1+\frac{1}{2}\|\dot{\bm{q}}\|\|\bm{s}\|$ with
$\varphi^*_f(\bm{\cdot})=\|\dot{\bm{q}}\|^2+\|\dot{\bm{q}}\|\|\bm{q}\|+\|\dot{\bm{q}}\|+\|\bm{q}\|+1$.
The initial conditions are: $\bm{q}(0)=[0.5;0.1;0.12]$,  $\dot{\bm{q}}(0)=[0;0;0]$, $\hat{\theta}(0)=0$.
The simulation results are shown in Figs. \ref{error3} and \ref{tau3}.
From Fig. \ref{error3}, it is observed that the outputs track the desired trajectory $\bm{q}_d$ and the tracking errors $\bm{e}$ converge to zero asymptotically.
The boundedness of the control input signals is illustrated in Fig. \ref{tau3}.

\begin{figure}[!htp]
  \centering
  \includegraphics[width=7.5cm]{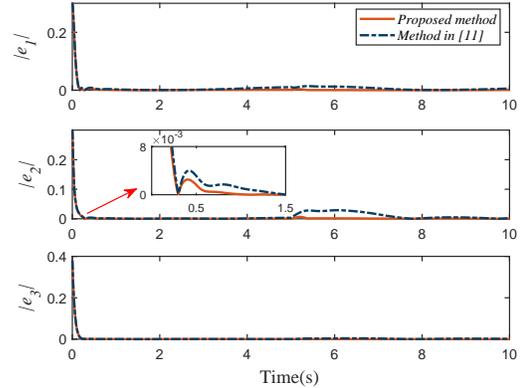}
  \caption{Comparison of tracking errors.}
  \label{cerror3}
\end{figure}

\begin{figure}[!htp]
  \centering
  \includegraphics[width=7.5cm]{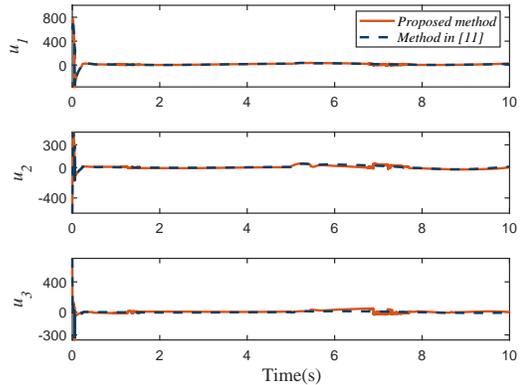}
  \caption{Comparison of control input signals.}
  \label{ctau3}
\end{figure}

To highlight the intriguing performance properties of the proposed method, we carry out fair comparisons with a traditional method developed in \cite{Song-Huang2016tac} with integrable function chosen as $\nu(t)=0.5$.
{Besides, owing to the control direction of the system in \cite{Song-Huang2016tac} has been assumed to be known, we use controller (\ref{ru2}) to compare with it.}
It can be seen from Figs. \ref{cerror3} and \ref{ctau3} that the proposed method exhibits stronger robustness when the undetectable fault occurs, yet the control inputs will chatter as $\nu(t)$ grows smaller.

\section{Conclusion}
In this paper, a robust adaptive tracking control method with controllability relaxation is proposed for a class of MIMO systems with unknown nonlinearities and unknown control directions.
By introducing some feasible auxiliary matrix, the strong controllability conditions for a large class of MIMO systems are relaxed and further extended to the case with unexpected intermittent actuator faults.
In addition, for time-varying input gain of an unknown sign (positive or negative), global asymptotic tracking control is achieved by embedding a novel Nussbaum function and certain positive integrable function in the control design.
Moreover, the control scheme obviates resorting to any linearization and approximation and enables automatic compensation of failed actuators without fault detection and diagnosis module, with the advantages of low-complexity structure and less-expensive computation.
Finally, application and simulation examples in robotic systems confirm the effectiveness of the proposed method. Extension such method to the more general pure-feedback MIMO systems represents a topic further work \cite{Huang-Song2020auto}.

\bibliographystyle{IEEEtran}
\bibliography{reference}

\end{document}